# Mn-substitution effects on MgB$_2$ superconductor

Sheng Xu[1], Yutaka Moritomo[2]*, Kenichi Kato[3] and 

[1]*Department of Crystalline Materials Science, Nagoya University, Nagoya 464-8603, Japan*
[2]*CIRSE, Nagoya University, Nagoya 464-8601, Japan*
[3]*JASRI, ???*

Effects of magnetic impurity on the superconductivity has been investigated in Mg$_{1-x}$Mn$_x$B$_2$. With increase of Mn concentration $x$, the lattice constant $c$ (perpendicular to the boron sheet) decreases at a rate of -1.4 % (= $\mathrm{d}\ln(c)/\mathrm{d}x$), while $a$ remains nearly unchanged. The transition temperature $T_\mathrm{c}$ steeply decreases with $x$ at a rate of $\mathrm{d}T_\mathrm{c}/\mathrm{d}x$ = - 159 K. These results suggest that superconducting state of the parent MgB$_2$ is amenable to the magnetic impurities, *i.e.*, Mn$^{2+}$.

KEYWORDS: MgB$_2$, chemical substitution

The recent discovery of the superconductivity in MgB$_2$ at $T_\mathrm{c}$ = 39 K[1] has stimulated world wide excitement. This is not only due to the simplicity in the chemical composition, the crystal structure and electronic structure, but also due to the its potentiality for application. Many researches are accumulating the details informations on the physical properties of the parent MgB$_2$. On the other hand, the chemical substitution is one of the powerful approach not only to reveal the nature of the parent material, but also to enhance the material potentiality by elevating the transition temperature $T_\mathrm{c}$. MgB$_2$ has a hexagonal structure (AlB$_2$-type; space group $P6/mmm$)[2] with alternating B- and Mg-sheets. The hexagonal network of the two-dimensional boron sheet governs the electronic structure near the Fermi level,[3–5] and hence is believed to be responsible for the superconductivity. Band structural calculations showed that the electronic structure is rather three-dimensional, making a sharp contrast with the two-dimensional electronic structure of the graphite intercalation compounds. Another structural feature of MgB$_2$ is that a large number isostructural compounds, such as, AlB$_2$, CrB$_2$ and MnB$_2$, exist. These structural features have motivated attempts to substitute Mg with Li,[6] Al[7] and Zn,[8] and B with C.[9]

In this Letter, we report effects of the substitution of the Mn$^{2+}$ ions for the Mg$^{2+}$ ions on the superconductivity of MgB$_2$. We have found a significant suppression of $T_\mathrm{c}$ with Mn concentration $x$, even though Mg$_{1-x}$Mn$_x$B$_2$ is isoelectrical to the parent MgB$_2$. Such a suppression of $T_\mathrm{c}$ due to the magnetic impurities, which locate outside of the boron sheets, perhaps reflects the three-dimensional electronic band structure of MgB$_2$.

The Mg$_{1-x}$Mn$_x$B$_2$ ($x$ =0.0, 0.01, 0.03, 0.05, 0.10 and 0.15) samples were synthesized by heating a stoichometric mixture of amorphous boron (98 %), magnesium powder (99.9 %) and Mn powder (99.9 %) at 900 °C for 2 hour. The powders are place in the Ta foil and heated in a flow of Ar/H$_2$5% gas. An x-ray powder patterns are measured at BL02B2 beamline at SPring-8. The samples were crushed into a fine powder and sealed in a 0.3mm $\phi$ quartz capillary, which gives a homogeneous intensity distribution in the Debye-Scherrer powder ring. Lattice constants were refined by the Rietveld analysis. The temperature dependence of magnetization $M$ was measured in a form of a lump of powders, in a Quantum Design PPMS magnetometer under an applied field of 10 Oe. The data were taken on heating after cooling down to the lowest temperature in zero field (ZFC).

First of all, let us show in Fig.1 the whole x-ray powder pattern (cross) of the parent MgB$_2$ at 300 K. The wavelength of the x-ray is $\approx$ 0.5 Å. To accurately determine the lattice constants, $a$ and $c$, we have analyzed the powder pattern with the RIETAN-2000 program[10] with AlB$_2$ structure ($P6/mmm$, No 191) with Mg at (0,0,0) and B at (1/3,2/3,1/2). Solid curve is result of the calculation. All the reflections observed up to 50° can be indexed with the AlB$_2$ structure (see the vertical line of Fig.1) The final refinements are satisfactory, in which $R_\mathrm{wp}$ (reliable factor based on the integrated intensity) are fairly reduced. The lattice constants are determined to be $a$ = 3.8200(9) Å and $c$ = 3.52166(9) Å. We further have measured the temperature variation of resistivity for the parent MgB$_2$ (not shown) by means of the four-probe method: the resistivity becomes zero below $T_\mathrm{C}$ = 38.5 K.

Figure 2 shows the whole x-ray powder pattern of the parent Mg$_{1-x}$Mn$_x$B$_2$ ($x$ =0.0, 0.05, 0.10 and 0.15) at 300 K. In all the compounds, sharp reflections from the AlB$_2$ structure are observed. At $x$ = 0.10 ad 0.15, however, several small impurity peaks are observed. These impurity peaks can be ascribed to MgB$_4$ (open triangles) and MgO (closed triangle). Inset shows the magnified patterns around the (002) reflection. The reflection shift toward the high-angle side with $x$, indicating decrease of the inter-plane B-B distance. This substitution effect is consistent with the lattice constants of MnB$_2$: $c$ is much shorter ($c$ = 3.0367(2) Å) in MnB$_2$ as compared with

* To whom correspondence should be addressed

MgB$_2$. We show in Fig.3 doping dependence of the lattice constants: (a) $a$ and (b) $c$. Lattice constants were refined by the Rietveld analysis with removing the impurity peaks. With increase $x$, $c$ decreases at a rate of -1.4 % (= dln$c$/d$x$), while $a$ remains nearly unchanged. This type of substitution effects on the lattice structure is analogous to Mg$_{1-x}$Al$_x$B$_2$,[7] but is in sharp contrast with the Li-doped[6] and C-doped MgB$_2$.[9] In the latter cases, the inter-plane B-B distance is essentially unchanged.

Figure 4 shows the temperature dependence of susceptibility $\chi$ for Mg$_{1-x}$Mn$_x$B$_2$. The data were taken under applied filed of 10 Oe on heating after cooling in zero field (ZFC). The samples at $x$ = 0.00, 0.01 and 0.03 show well-defined one-step transitions as well as large shielding fraction before correction of demagnetization, indicating that the superconductivity is of bulk nature. The transition temperature $T_c$, defined by the intersection of the extrapolated lines below and above the transition, decreases with increase of $x$ at a rate of d$T_c$/d$x$ = - 159 K (see downward arrows). The transition, however, becomes blurred with further increase of $x$ beyond 0.05. Such a blurred feature of the transition may be ascribed to inhomogeneous distribution of the Mn ions and resultant distribution of $T_c$. The $c$-coefficient of $T_c$, - dln$T_c$/d$c$, is estimated to be -83 Å$^{-1}$.

Finally, let us compare the present substitution effect on $T_c$ with the substitution effects of the other elements. We show in Fig.5 the variation of $T_c$ for Mg$_{1-x}$Mn$_x$B$_2$ against $x$, together with the data for Mg$_{1-x}$Al$_x$B$_2$ (cited from Ref.[7]) and Mg$_{1-x}$Zn$_x$B$_2$ (cited from Ref.[8]). The suppression of $T_c$ is much steeper than the case of the Al- and Zn-substitutions. In addition, the $x$-coefficient of $T_c$, d$T_c$/d$x$ = - 159 K, is much larger than that of the MgB$_{2-x}$C$_x$ system (d$T_c$/d$x$ = - 57 K[9]). Such a large $x$-coefficient of $T_c$ cannot be ascribed to the lattice structural change, because absolute magnitude of the $x$-coefficient of $c$, dln$c$/d$x$ = -1.4 %, is comparable to those for Mg$_{1-x}$Zn$_x$B$_2$ (dln$a$/d$x$ = + 1.7 %, dln$c$/d$x$ = + 2.0 %).[8] Here, note that the Mn$^{2+}$ ions has five $d$-electrons, making a sharp contrast with the non-magnetic Zn$^{2+}$ or Al$^{3+}$ ions. Our experimental result suggests that the superconducting state of the parent MgB$_2$ is amenable to the spin impurity, even though the impurities locate at the Mg-sheet outside of the boron layers.

In summary, effects of magnetic impurity on the superconductivity has been investigated in Mg$_{1-x}$Mn$_x$B$_2$. We have found a significant suppression of $T_c$ with Mn concentration $x$, the magnetic impurities, *i.e.*, Mn$^{2+}$, locate outside of the boron sheets. This observation perhaps reflects three-dimensional electronic band structure of MgB$_2$. A more elaborated investigation on the chemical substitution effects on $T_c$ is necessary to reveal the nature of MgB$_2$ superconductor.

**Acknowledgements**

This work was supported by a Grant-in-Aid for Scientific Research from the Ministry of Education, Science, Sports and Culture. The synchrotron power experiments were performed at the SPring-8 BL02B2 with approval of the Japan Synchrotron Radiation Research Institute (JASRI).

Fig. 1. The whole x-ray powder pattern (crosses) at 300 K for MgB$_2$. Solid curve is the results of the Rietveld refinement with hexagonal structure ($P6/mmm$).

Fig. 2. The whole x-ray powder pattern at 300 K for Mg$_{1-x}$Mn$_x$B$_2$. Open triangle and closed triangle represent MgB$_4$ and MgO impurities. Insets show the magnified patterns around the (002) reflection.

Fig. 3. Doping dependence of lattice constant, (a) $a$ and (b) $c$, for Mg$_{1-x}$Mn$_x$B$_2$.

Fig. 4. Temperature dependence of susceptibility $\chi$ for Mg$_{1-x}$Mn$_x$B$_2$. The data were taken under applied filed of 10 Oe on heating after cooling in zero field (ZFC). Downward arrows indicate the transition temperatures $T_c$.

Fig. 5. Doping dependence of $T_c$ for Mg$_{1-x}$Mn$_x$B$_2$. Broken lines are data for Mg$_{1-x}$Al$_x$B$_2$ (cited from Ref.7) and Mg$_{1-x}$Zn$_x$B$_2$ (cited from Ref.8).

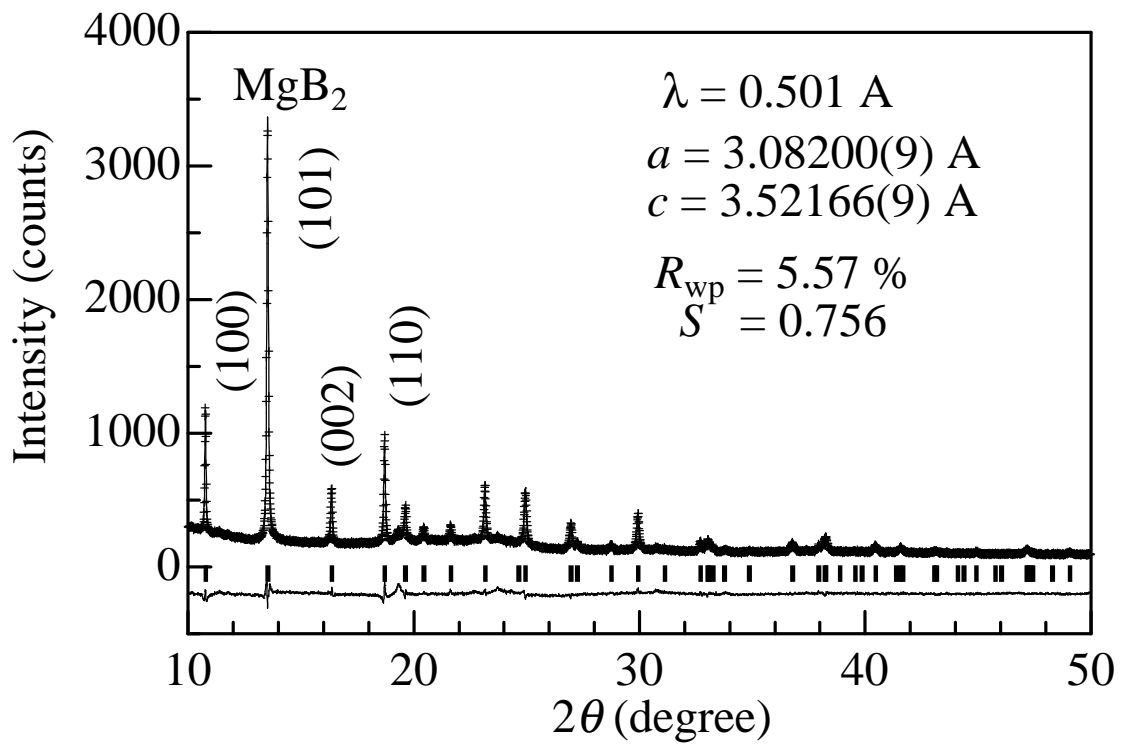

Fig.1: Sh. Xu *et al.*; JPSJ

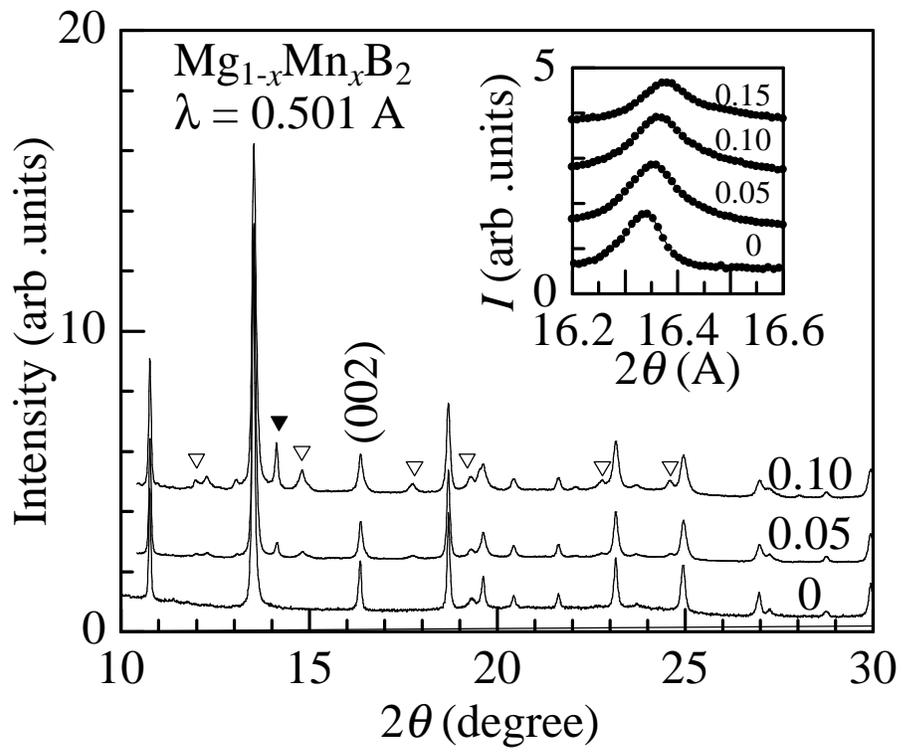

Fig.2: Sh. Xu, *et al.*, JPSJ

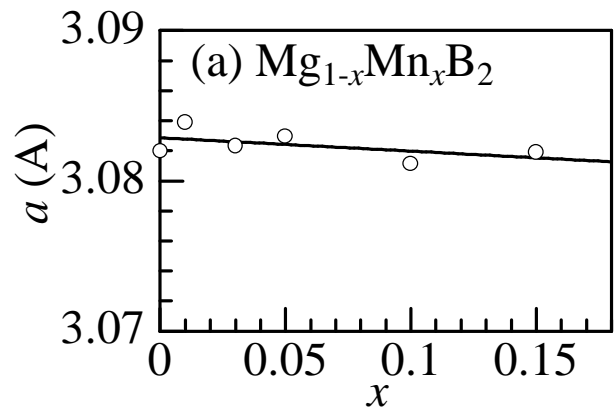
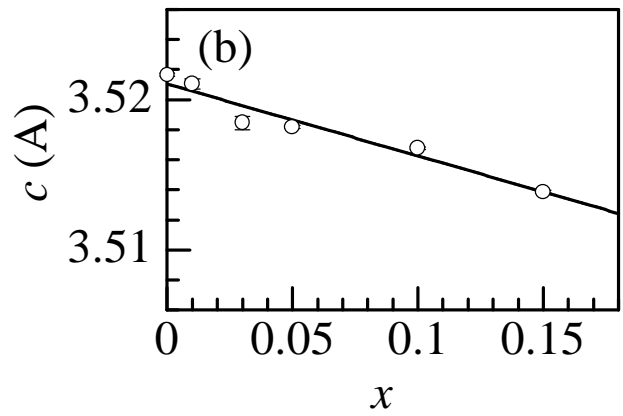

Fig.3: Sh. Xu *et al.*, JPSJ

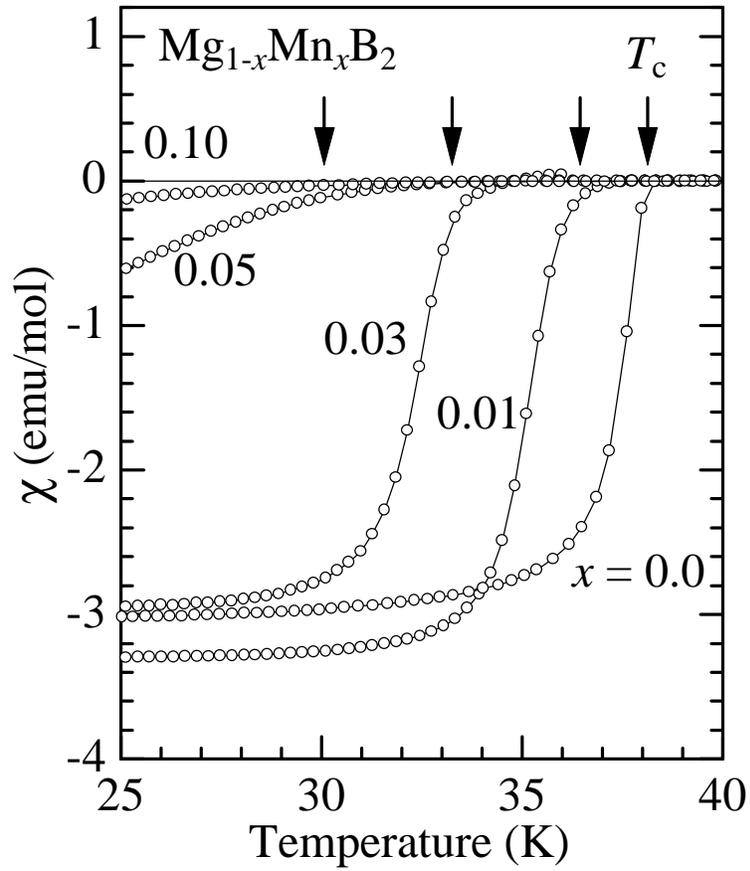

Fig.4: Sh, Xu, *et al.*, JPSJ

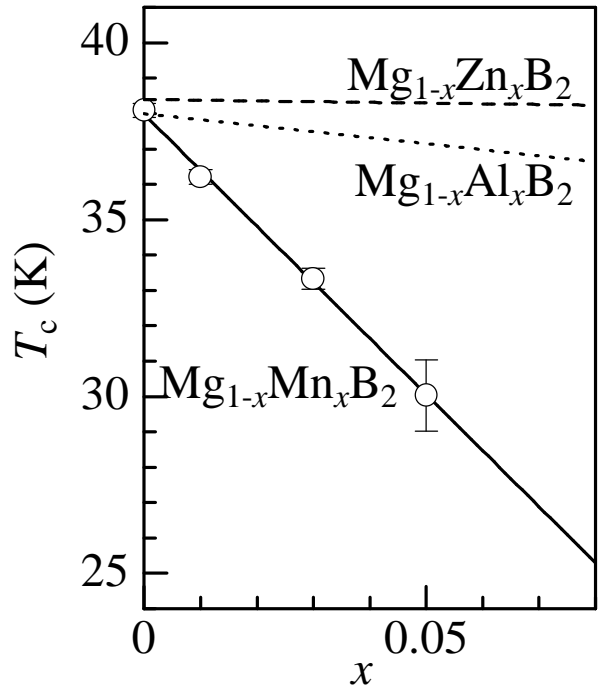

Fig.5: Sh, Xu, *et al.%,; JPSJ*